\documentclass[%
 reprint,
superscriptaddress,
 amsmath,amssymb,
 aps,
longbibliography,
]{revtex4-1}

\usepackage{graphicx}
\usepackage{dcolumn}
\usepackage{bm}

\begin{document}

\preprint{APS/123-QED}

\title{Dark-soliton molecules in an exciton-polariton superfluid}

\author{Anne Ma\^{i}tre}
\email{anne.maitre@lkb.upmc.fr}
\affiliation{Laboratoire Kastler Brossel, Sorbonne Universit\'{e}, CNRS, ENS-Universit\'{e} PSL, Coll\`{e}ge de France, Paris 75005, France}%

\author{Giovanni Lerario}%
\affiliation{Laboratoire Kastler Brossel, Sorbonne Universit\'{e}, CNRS, ENS-Universit\'{e} PSL, Coll\`{e}ge de France, Paris 75005, France}%
\affiliation{CNR NANOTEC, Istituto di Nanotecnologia, via Monteroni, 73100 Lecce, Italy}

\author{Adri\`{a} Medeiros}%
\affiliation{Laboratoire Kastler Brossel, Sorbonne Universit\'{e}, CNRS, ENS-Universit\'{e} PSL, Coll\`{e}ge de France, Paris 75005, France}%

\author{Ferdinand Claude}%
\affiliation{Laboratoire Kastler Brossel, Sorbonne Universit\'{e}, CNRS, ENS-Universit\'{e} PSL, Coll\`{e}ge de France, Paris 75005, France}%

\author{Quentin Glorieux}%
\affiliation{Laboratoire Kastler Brossel, Sorbonne Universit\'{e}, CNRS, ENS-Universit\'{e} PSL, Coll\`{e}ge de France, Paris 75005, France}%

\author{Elisabeth Giacobino}%
\affiliation{Laboratoire Kastler Brossel, Sorbonne Universit\'{e}, CNRS, ENS-Universit\'{e} PSL, Coll\`{e}ge de France, Paris 75005, France}%

\author{Simon Pigeon}%
\affiliation{Laboratoire Kastler Brossel, Sorbonne Universit\'{e}, CNRS, ENS-Universit\'{e} PSL, Coll\`{e}ge de France, Paris 75005, France}%

\author{Alberto Bramati}%
\affiliation{Laboratoire Kastler Brossel, Sorbonne Universit\'{e}, CNRS, ENS-Universit\'{e} PSL, Coll\`{e}ge de France, Paris 75005, France}%


\date{\today}

\begin{abstract}

The general theory of dark solitons relies on repulsive interactions and therefore predicts the impossibility to form dark-soliton bound states.
One important exception to this prediction is the observation of bound solitons in non-local nonlinear media \cite{Dreischuh2006}.
Here, we report that exciton-polariton superfluids can also sustain dark-soliton molecules although the interactions are fully local.
With a novel all optical technique, we create two dark solitons and bind them to each other to form an unconventional dark-soliton molecule.
We demonstrate that the stability of this structure and the separation distance between two dark-solitons is tightly connected to the driven-dissipative nature of the polariton fluid.


\end{abstract}

\maketitle


\section{\label{sec:level1}Introduction}
In semiconductor microcavities, the strong coupling between quantum well excitons and cavity photons gives rise to two quasi-particles with distinct energies: the upper and lower polaritons. They combine properties from light and matter, which makes them behave like weakly interacting composite bosons. 
Their photonic part gives them a light effective mass and the possibility to be excited and detected optically, while their excitonic part is responsible for their interactions.
Exciton-polaritons have revealed themselves to be an interesting platform to investigate bidimensional bosonic quantum fluids. Indeed, several effects have been observed in such systems: nonequilibium phase transition similar to Bose-Einstein Condensation \cite{Deng2002, Kasprzak2006a, Deng2010}, superfluidity \cite{Amo2009}, or the Josephson effect \cite{Sarchi2008, Shelykh2008}. 

Phenomena linked to the hydrodynamics of quantum fluid were also reported, as the nucleation of quantized vortices \cite{Sanvitto2011, Nardin2011, Pigeon2011, Boulier2018, Boulier2016a} or the formation of dark solitons \cite{Amo2011, Grosso}.
These solitons are stable collective excitations resulting from the balance between the diffraction and the fluid repulsive interactions.
They are characterized by a density dip coinciding with a phase jump, while their shape is maintained all along their propagation. 
They were observed in several nonlinear systems, such as cold atoms condensates \cite{Frantzeskakis2010, Denschlag2000}, thin magnetic films \cite{Chen1993, Shinjo2000}, liquid Helium \cite{Williams1999} or optical fibers \cite{Weiner1988}.
Their interactions were also studied in different systems \cite{Foursa1996,Stellmer2008}.
Differently from bright solitons which may attract and form molecules, dark solitons interactions are predicted to always be repulsive so that bound states of dark solitons cannot form in nonlinear local media \cite{Frantzeskakis_2010,zhao1989interactions,PhysRevLett.66.1583}.
Interestingly, in nonlinear media with non-local interactions, such attractive behaviour between dark solitons has been predicted \cite{Nikolov04,Kong2010} and experimentally observed in a thermo-optic medium \cite{Dreischuh2006}. 

In this manuscript, we present experimental and numerical results in striking contrast with these predictions.
We demonstrate the all-optical creation and binding of dark solitons, forming a dark soliton-molecule \cite{martinez2011creation} in a purely local medium.
This unconventional behavior for dark solitons, which strongly deviates from equilibrium atomic quantum fluids, is attributed to the driven-dissipative nature of the polariton quantum fluid, and we show that the  specific equilibrium separation distance between the bound solitons is a direct consequence of the finite polariton lifetime.

\section{Bistability enhancement of polariton free propagation}
The resonant pumping of exciton-polaritons consists in injecting photons into the system at a specific energy and in-plane wavevector of the lower polariton branch. 
At high driving intensity, it results in the creation of a polariton fluid whose phase coincides with the pump one. In this strongly driven regime, superfluid effects can be studied, like viscus-free flow \cite{Amo2009}. However, the presence of the driving field prevents the free evolution of the fluid properties \cite{Pigeon2011}. To overcome this constraint, spatial or temporal engineering of the excitation beam has been used, leading to the observation of quantized vortices or dark solitons pairs \cite{Amo2011, Sanvitto2011, Pigeon2011, Boulier2018, Boulier2016a, Nardin2011} in the driving-free region, but with a propagation distance limited by the short lifetime of the polaritons. 

The introduction of a small positive energy detuning between the driving field and the lower polariton branch induces an optical bistability \cite{Baas2004}, offering two stable states for the system (figure \ref{fig:setup}a): a linear regime with a low polariton density and negligible interactions between particles, while the high density regime presents strong interactions and corresponds to the superfluid regime \cite{Carusotto2004a}.
The system exhibits a hysteresis cycle, i.e. a range of input intensities where the two states are accessible, called the bistable regime in the following and highlighted in white in figure~\ref{fig:setup}a.

Recently, a theoretical proposal by Pigeon \textit{et al.} \cite{Pigeon2017} suggested to exploit the optical bistability to  compensate the system losses without inhibiting the fluid spontaneous dynamic. 
Supporting the fluid with an intensity within the bistable regime releases the constraint imposed by the quasi-resonant driving field and makes possible the spontaneous formation of turbulence sustained for macroscopic propagation distances \cite{lerario2020vortex}.

The typical technique to generate fluid turbulence, vortices and solitons in exciton-polaritons systems is to send the fluid towards an obstacle \cite{Amo2011, Sanvitto2011, Pigeon2011, Nardin2011}, but this configuration constrains the study of these excitations to conditions where they can form spontaneously. 
To overcome this limitation, we present here an original method where the solitons are directly imprinted with the use of a Spatial Light Modulator (SLM). Combining beam shaping techniques for direct imprinting and the bistable support mechanism \cite{Pigeon2017} for long propagation length, we report the controlled generation of dark soliton bound states and their free propagation over macroscopic distances.
\\

\section{Experimental setup}

The sample used in the present work is a planar microcavity which contains three $In_{0.04}Ga_{0.96}As$ quantum wells at the antinodes of the cavity electromagnetic field. The GaAs cavity has a length of \(2\lambda/n_{c}\), with \(\lambda = 835\) nm the excitonic resonance and $n_{c}$ = 3.54 the cavity GaAs substrate optical index; the two Bragg mirrors consist of 21 and 24 GaAs-AlAs pairs. The polariton lifetime is extracted from time resolved measurements to be 10 ps while the Rabi splitting is 5.1 meV and the linewidth 0.07 meV.

The polariton fluid is obtained by quasi-resonant pumping using a continuous wave single-mode Ti-Sapphire laser at a frequency slightly blue detuned from the lower polariton branch resonance (typically an energy detuning $\Delta E = 0.20$ meV).

In order to imprint a particular phase profile on the fluid, one needs to excite the system above the bistability regime to fix the polariton phase properties to the pump ones. On the other hand, a free propagation of collective excitations requires a bistable fluid, in which the phase is free to evolve. 
Therefore, the implementation of such imprinting mechanism requires an intensity gradient, providing a region of high intensity where the phase pattern is imprinted into the fluid, and a region of lower intensity, bistable, where the driving phase is flat and where the imprinted modulations are able to propagate freely.

\begin{figure}[t]
\includegraphics[width=0.49\textwidth]{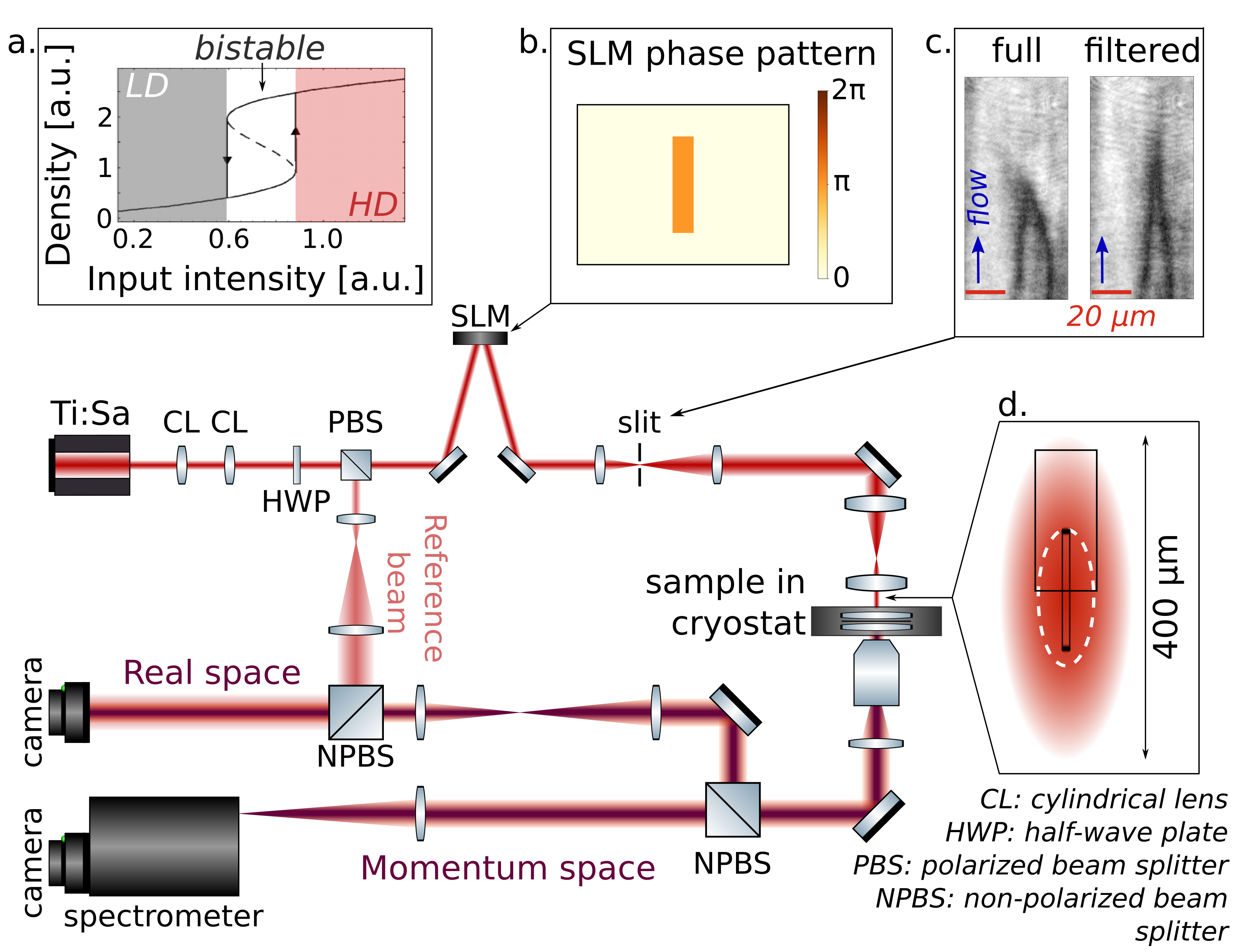}
\caption{\label{fig:setup} Experimental setup. Inset (a) shows the theoretical bistability curve resulting from the quasi-resonant pumping. The grey region corresponds to the low density regime (LD) while the red one has much higher polariton density (HD). The white part delimits the bistable hysteresis cycle.\\ 
The phase profile of the excitation beam is customized with the SLM, as plotted in inset (b). The beam is then spatially filtered to smooth the phase jump: inset (c) illustrates how the opening of the filtering slit influences the phase gradient in the transverse direction.
The inset (d) shows the shape of the pump imaged into the sample: the white dashed line corresponds to the upper bistability frontier, while the black rectangle indicates the detection field of view. The real space image gives access to the density and phase maps (interfering with a reference beam), whereas the conditions of the driving field are extracted from the momentum space detection.}
\end{figure}

This configuration is reached by using the gaussian shape of the exciting beam (see figure \ref{fig:setup}d). By adjusting the intensity of the pump, the central part of the illuminated area is placed above the bistability limit, i.e. in the red part of figure \ref{fig:setup}a, imprinting its phase to the fluid. 
The region where the local pump intensity is above the bistability cycle is delimited by the white dashed line:  outside this region, the pump has a lower intensity and induces a bistable fluid, whose phase is free to evolve.
The detection field of view (black rectangle) is placed in order to observe the propagation of the imprinted excitation.
The exciting beam on the sample has a vertical dimension of 400 microns, which induces an intensity difference of more than 10 \% between the center of the spot and the region 150 $\mu$m above it. 

\section{All optical imprinting of parallel dark solitons}

The phase pattern (figure \ref{fig:setup}b) is imprinted on the pump beam using an SLM, whose surface is imaged onto the sample.
This allows for a direct control of its shape and position, so that it can be placed at the limit of the bistable regime of the fluid.
We design a rectangular shape where the phase is $\pi$ - shifted with respect to the outer beam (figure \ref{fig:setup}b).
A spatial filtering is realized in the Fourier plan before imaging the SLM plan onto the sample: high spatial frequencies components are cut in the direction of the flow in order to smooth the phase jump experienced by the flow at the end of the phase pattern.
The cutoff directly influences the shape of the top part of the solitons imprinted on the fluid, as a smoothing of the phase modulation induces an elongation of the solitons, illustrated in inset c of figure \ref{fig:setup}. When the slit is fully opened, the phase jump along the direction of the flow is sharp, leading to a pair of imprinted solitons closing on a short distance. However, when the slit aperture is reduced, the imprinted solitons extend along the flow, gradually approaching to each other. Such filtering reduces their respective lateral velocity and favors their parallel propagation.

\begin{figure}[b]
\includegraphics[width=0.49\textwidth]{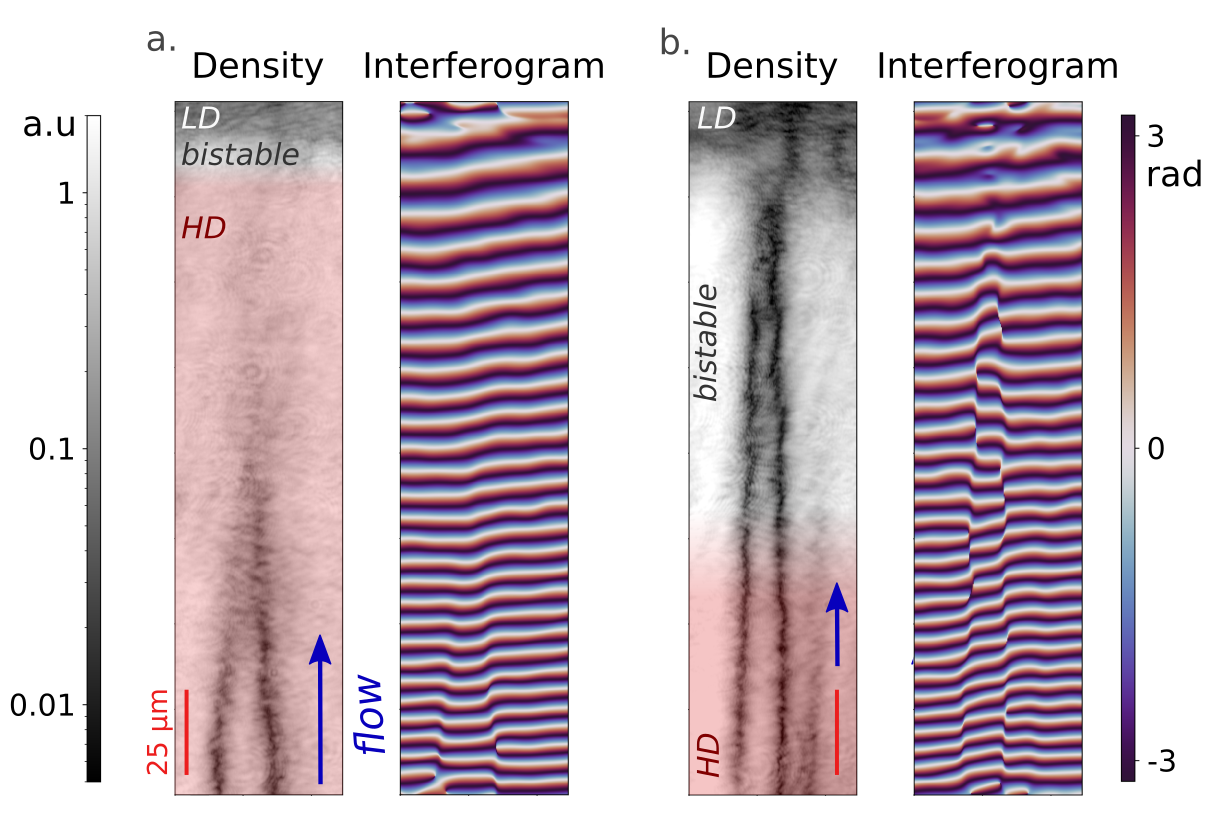}
\caption{\label{fig:solparr} Density and phase of imprinted solitons propagating parallel. 
(a) Density (logarithmic scale) and phase at high power, where almost all the pumped area is above the bistability range, and therefore gets the phase profile corresponding to the driving one. The top of the picture corresponds to the limit of the excitation: one can see a thin bistable region just before the density of polariton abruptly drops to the low density regime. 
(b) At lower intensity: the white area corresponds to the bistable part of the fluid where the solitons propagate freely, until they reach the lower part of the bistability range. The energy detuning $\Delta E$ is 0.13 meV, the fluid velocity is 1.35 $\mu$m/ps, c$_{s}$ = 0.31 $\mu$m/ps and m$^{*} = 2.20 \cdot 10^{-34} $ kg.}
\end{figure}

To be sustained by the fluid, the solitons need supersonic conditions \cite{Kamchatnov2008,Amo2011}: those are achieved by choosing the appropriate wavevector and energy detuning of the driving field.
The sound velocity c$_{s}$ depends directly on the energy detuning $ \Delta E $ between the laser and the lower polariton branch:
 c$_{s} \approx \sqrt{ \dfrac{\Delta E}{m^{*}} }$, with m$^{*}$ the polariton effective mass. The speed of the fluid is linked to the in-plane wavevector as it is the derivative of the dispersion: 
$ v_{f} = \dfrac{\partial \omega_{LP}(\mathbf{k}_{\parallel})}{\partial \mathbf{k}_{\parallel}} $ with $\mathbf{k}_{\parallel}$ the driving field in-plane wavevector.

\begin{figure}[t]
\includegraphics[width=0.49\textwidth]{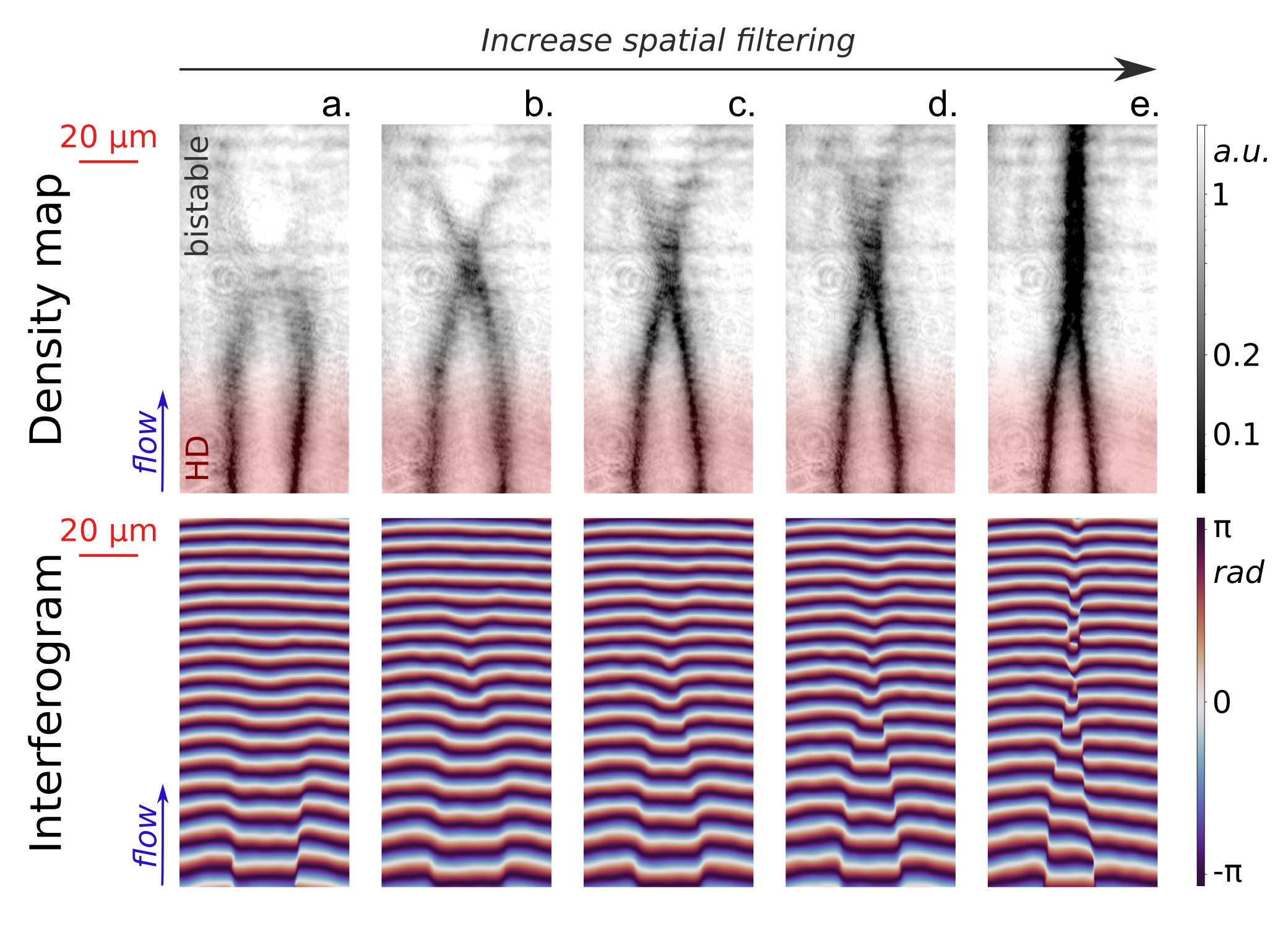}
\caption{\label{fig:setslit} Set of solitons impression with different spatial filtering. On (a), the phase jump is not filtered: on the fluid, it is blurred by the flow but the solitons do not propagate on the bistable region. From picture (b) to (d), the filtering of the spatial components is gradually increased. The solitons start to propagate, but their respective lateral velocities remain too high and they bounce on each other, vanishing along their propagation. On (e), the spatial components are filtered enough so that they can align and propagate parallel, remaining dark solitons. For this set, the detuning between the laser and the lower polariton branch $\Delta E$ is 0.11 meV, the flow speed is $v_{f} = 0.66$ $\mu$m/ps, the speed of sound $c_{s} = 0.42$ $\mu$m/ps and the effective polariton mass $m^{*} = 1.04 \cdot 10^{-34}$ kg
}
\end{figure}

In figure \ref{fig:solparr}, we illustrate the crucial role of the bistable regime with respect to the solitons behaviour. The polariton flow is from bottom to top, and the red highlighted areas indicate the regions of the fluid above the bistability cycle, where the phase is fixed by the driving field. In figure \ref{fig:solparr}a, we use a high laser intensity so that the main part of the pumped region reproduces the pump pattern: the density and phase modulation are indeed only visible in the bottom part of the picture. 
The interferograms are obtained using homodyne interferometry with a reference beam, extracted from the pump beam before any shaping through the SLM. The interference images are then Fourier filtered to enhance the phase pattern.
The high power images (fig. \ref{fig:solparr}a) are used as a reference to locate the position of the imprinted solitons.
Lowering gradually the intensity of the driving field, the bistable region of the fluid expands from the edge of the locked area towards the center of the driving beam, eventually reaching the imprinted solitons. The phase of the fluid then spontaneously readjusts to let the soliton pair propagate with the flow (figure \ref{fig:solparr}b), despite the phase mismatch with the driving field.
We can observe that the solitons remain dark with a phase jump of $\pi$ and parallel, as long as they are sustained by the fluid. As they propagate on the outer part of the excitation spot, the intensity of the pump is slowly decreasing until it reaches the threshold of the lower part of the bistability loop; the fluid then jumps to the low density regime where no solitons can be sustained anymore.

\section{Dark-soliton bound states}

The parallel propagation of the solitons assures them to stay dark and to maintain their phase jump. However, this configuration is not exclusive: one can find some situations where the solitons reflect on each other, leading them to open and vanish along their propagation (see figure \ref{fig:setslit}). In that case they become grey solitons, as their phase jump is not  $\pi $ anymore, and can be described as Bloch domain walls. 
In particular, a sensitive parameter is the spatial filtering of the phase jump described previously, as shown in figure \ref{fig:setslit}. The opening of the slit in the Fourier space of the SLM directly influences the shape of the solitons on the fluid: on figure \ref{fig:setslit}a, the slit is open and the phase jump is well defined when entering the cavity; it only blurs on the fluid because of the flow. In this case, the solitons do not propagate in the bistable region.
From figure \ref{fig:setslit}b to figure \ref{fig:setslit}d, the slit is gradually closed, leading to a higher spatial filtering and a longer propagation of the solitons. Yet, their respective lateral velocity is still too high for them to align: they bounce on each other, reflect, and vanish along their propagation. They reflect with the same angle with which they meet, respectively 25$^{\circ}$, 17$^{\circ}$ and 13$^{\circ}$ for figure \ref{fig:setslit}b, \ref{fig:setslit}c and \ref{fig:setslit}d. Finally, in figure \ref{fig:setslit}e, they meet with an even smaller angle (11$^{\circ}$) and align, propagating parallel for a hundred microns. The pictures of figure \ref{fig:setslit} correspond to a fluid velocity $v_{f} = 0.66$ $\mu$m/ps, a speed of sound $c_{s} = 0.42$ $\mu$m/ps, an effective polariton mass $m^{*} = 1.04 \cdot 10^{-34}$ kg and an energy detuning $\Delta E = 0.11$ meV. The flow is from bottom to top and the imprinting is done in the red region.

\begin{figure}[t]
\includegraphics[width=0.49\textwidth]{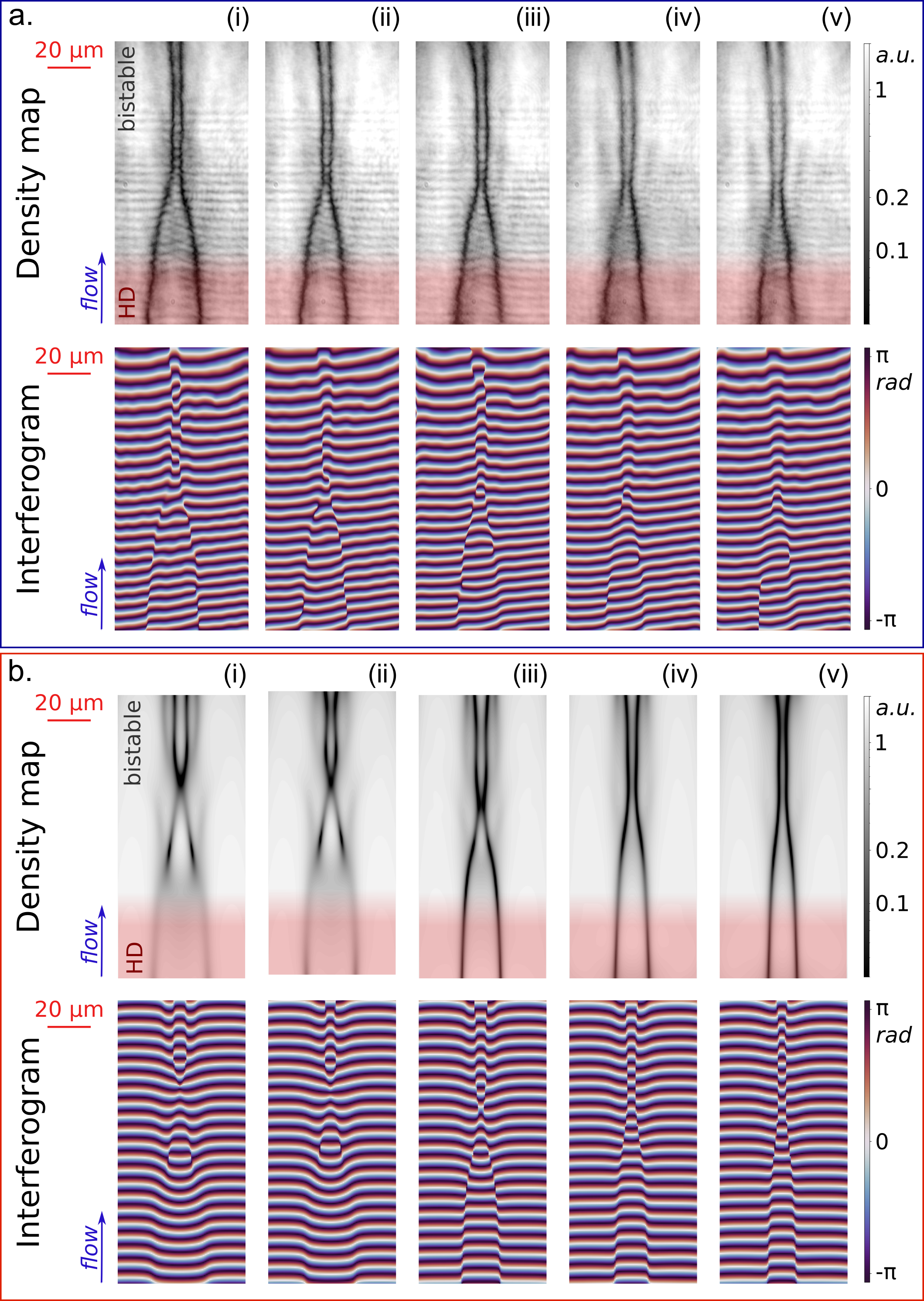}
\caption{\label{fig:setdist} Set of solitons aligning to each others for different imprinting distances. The red area indicates the imprinting region of the fluid. The fluid flow is from bottom to top.
(a) Experimental data, with the density map on top and the corresponding phase map below. Solitons get closer together along their propagation, until reaching a minimal distance from which they align parallel.
For this set, the blue shift is 0.34 meV, for a fluid velocity of 1.00 $\mu $m/ps, a speed of sound of 0.67 $\mu$m/ps, and an effective polariton mass of $ m^{*}=1.24 \cdot 10^{-34}$ kg.
(b) Numerical simulations which show the same behaviour: as soon as the driving intensity is reduced enough to reach the bistable regime, imprinted solitons enter a transition region where they get closer, until a minimal distance from which they align.
}
\end{figure}

In order to see the effect of the soliton-soliton interactions, we tune the imprinted separation distance and we observe that the freely propagating solitons tend to get closer to one another, before aligning at a specific distance, independently of the initial one.
Figure \ref{fig:setdist}a shows a set of measurements where the distance between the imprinted solitons vary: from 24 $\mu$m for figure \ref{fig:setdist}a(i) to 15 $\mu$m for figure \ref{fig:setdist}a(v). The fluid flow is still from bottom to top and the solitons imprinted in the red highlighted region. The fluid velocity is 1.00 $\mu$m/ps, $c_{s}$ = 0.67 $\mu$m/ps, the effective polariton mass of  m$^{*}=1.24\cdot 10^{-34}$ kg and the energy detuning $ \Delta E = 0.34 $ meV. Along their propagation, they get closer to one another, until reaching an equilibrium separation distance at which they align and propagate parallel for more than 50 $\mu$m. 
It is remarkable that the phase jump along the solitons line remains close to its imprinted value $\pi$.

Simulations based on a split-step method considering exciton-cavity photon basis \cite{Pigeon2011} are presented in panel b of figure \ref{fig:setdist}. The driving profile corresponds to a large Gaussian beam centered on the lower part of the image. It creates a slightly decreasing driving intensity from bottom to top.
In the high intensity region (in red in figure \ref{fig:setdist}) a $\pi$ phase jump is imprinted and the effect of the slit is reproduced smoothing the orthogonal phase jump, accordingly to the experiment.
The density and interferogram patterns presented correspond to steady state solutions. They reproduce accurately the experimental results, with an alignment of the solitons at a specific distance independently of their initial ones.

This convergence to an equilibrium configuration is analysed in figure \ref{fig:dprop}a. We show how the distance between the solitons evolves along their propagation, for different initial separation distances, identified by the different color lines.
After a transition region where the solitons get closer to one another, they all reach the same value, of $d_{sep}=4.8$ $\mu$m in this case. The separation distances are extracted from the images by fitting the transverse solitons profile, plotted in the inset of figure \ref{fig:dprop}a: 
\( |\psi (x)|^{2} = \tanh^{2} \Big( \frac{x-d_{sep}/2}{A} \Big) \tanh^{2} \Big( \frac{x+d_{sep}/2}{A} \Big) \) \cite{Pitaevskii2003a}, 
where A corresponds to the full width at half maximum of the solitons.

\section{Driven-dissipative stabilisation of dark-soliton molecules}
This parallel alignment is in striking contrast with the conventional behavior of dark solitons.
Far from repelling each other as expected \cite{Frantzeskakis2010} and observed \cite{Amo2011}, in the present situation they get closer to one another before reaching an equilibrium position.

In the region where the solitons propagate freely, the fluid between the solitons is out of phase with respect to the driving field. Consequently, the region of space remains effectively un-pumped, and is only populated by tunneling through the solitons.
A strong gradient of density is thus imposed across the solitons, between the outer region in phase with the driving and the out-of-phase inner region. The resulting pressure pushes the solitons toward each other, reducing the region in opposition of phase with the driving, until they feel the standard repulsion between them.
For well-adjusted parameters, the solitons do not annihilate each other and propagate, at the fixed distance where the driving pressure and their repulsive interaction equilibrate. 

\begin{figure}[t]
\includegraphics[width=0.45\textwidth]{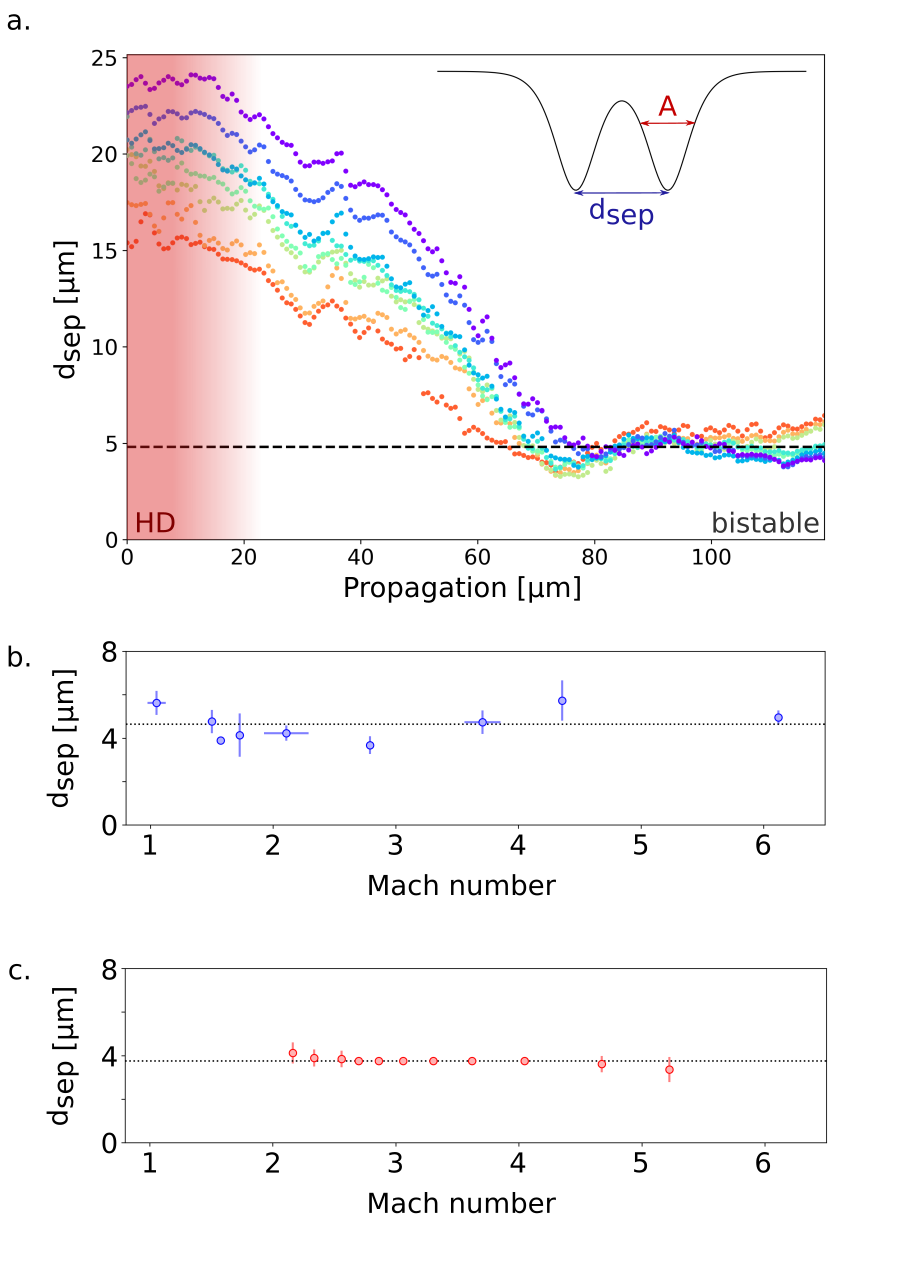}
\caption{\label{fig:dprop} 
(a) Evolution of the distance between the solitons along their propagation for different initial imprinting distances. Each color line corresponds to one picture with a different imprinted separation distance. Solitons tend to align at an equilibrium distance from each other. The energy shift $\Delta E = 0.34$ meV, the fluid velocity is 1.00 $\mu $m/ps, c$_{s} $ = 0.67 $\mu$m/ps and m$^{*} = 1.24 \cdot 10^{-34}$ kg. The black dashed line illustrates the equilibrium separation distance at 4.8 $\mu$m. The distances are extracted from the fit of the transverse soliton profile, shown in the inset.
(b) Summary of the experimental equilibrium separation distances for different hydrodynamic conditions, as a function of the Mach number. It stays constant around 4.75 $\mu$m. The parameters of each set are listed in table \ref{tab:setparam}.
(c) Numerical simulations are done by tuning only the energy shift $\Delta E$, and keeping constant the fluid velocity and the effective mass as the experimental ones.
}
\end{figure}

\begingroup
\begin{table}[b]

\begin{ruledtabular}
\begin{tabular}{ccccc}
\textrm{$\Delta E$ [meV]}&
\textrm{$c_{s}$ [$\mu$m/ps]}&
\textrm{$m^{*}$ [kg]}&
\textrm{$v_{f}$ [$\mu$m/ps]}&
\textrm{M}\\
\colrule

0.24 & 0.60 & 1.05$\cdot 10^{-34}$ & 0.62 & 1.05 \\
0.34 & 0.67 & 1.24$\cdot 10^{-34}$ & 1.00 & 1.50 \\
0.11 & 0.42 & 1.04$\cdot 10^{-34}$ & 0.66 & 1.57 \\
0.33 & 0.60 & 1.47$\cdot 10^{-34}$ & 1.03 & 1.73 \\
0.08 & 0.33 & 1.11$\cdot 10^{-34}$ & 0.70 & 2.11 \\
0.22 & 0.42 & 1.95$\cdot 10^{-34}$ & 1.17 & 2.79 \\
0.16 & 0.39 & 1.75$\cdot 10^{-34}$ & 1.43 & 3.71 \\
0.07 & 0.31 & 1.25$\cdot 10^{-34}$ & 1.33 & 4.36 \\
0.06 & 0.23 & 1.84$\cdot 10^{-34}$ & 1.42 & 6.12 \\

\end{tabular}
\end{ruledtabular}
\caption{\label{tab:setparam}%
Parameters of the experimental sets plotted in figure \ref{fig:dprop}b.
}
\end{table}
\endgroup

The parallel alignment of the solitons and their separation distance are expected to be very sensitive to the hydrodynamic conditions of the fluid. In particular the characteristic length of the system, the healing length $ \xi = \dfrac{\hbar}{\sqrt{2m^{*} \Delta E}} $, which for the typical experimental conditions of the set presented in figure \ref{fig:setdist} is 1.17 $\mu$m, should play a crucial role by setting the minimal achievable separation distance. The experiment was therefore performed in different hydrodynamic conditions in order to check this behaviour.
The detuning $ \Delta E$ between the laser and the lower polariton branch can easily be changed, however it induces a change of the bistability cycle and therefore perturbs the alignment of the solitons. This can be compensated by changing the in-plane wavevector - and thus the speed of the fluid - to find an other configuration favourable to the solitons parallel propagation.
The effective mass of the polaritons $m^{*}$ depends on the detuning between the bare exciton energy and the cavity photon one : it can be tuned experimentally by changing the working point on the sample, since it has a small wedge between the two Bragg mirrors. 
Severals sets of experiments have been realised; the parameters are listed in table \ref{tab:setparam} and the results are summarized in figure \ref{fig:dprop}b. Each point corresponds to the average equilibrium distance of a set of free propagating solitons, in the same hydrodynamic conditions but imprinted with different initial separation distances. 
 The equilibrium distances obtained for each set of measurements are plotted here as a function of the Mach number $M=v_f/c_s$ : quite surprisingly, no significant changes appear in the equilibrium separation distance of the solitons, no matter the considered experimental conditions. The equilibrium distance is therefore independent from the healing length of the fluid.

As previously explained, experimentally those parameters can hardly be studied independently, however numerically they can be separately tuned. Figure \ref{fig:dprop}c presents the calculated separation distance of the soliton pairs as a function of the Mach number. In the simulations, only the detuning between the pump and the lower polariton branch $\Delta E$ is changed; the effective mass and the speed of the fluid are kept constant, accordingly to the experimental conditions. The range of authorized Mach number is therefore smaller than the one observed in the experiments. However, we clearly see that the separation distance is mostly unaffected by the change of the Mach number. 
The error bars come from the small breathing of the soliton pair, similar to what was predicted in Ref. \cite{Kong2010,Koniakhin2019}. Such effects are too small to be accurately resolved experimentally in the present conditions. From figures \ref{fig:dprop}b and \ref{fig:dprop}c, we can conclude that the separation distance is independent of the fluid hydrodynamics properties.

\begin{figure}[t]
\includegraphics[width=0.45\textwidth]{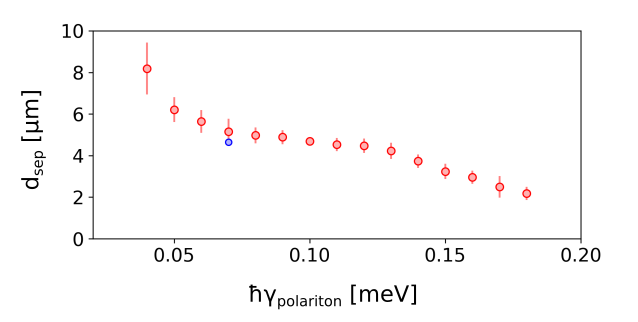}
\caption{\label{fig:dsepgamma} 
Numerical simulations of the equilibrium separation distance as a function of the decay rate of the polariton (red dots). The blue dot corresponds to the experimental value for our sample. The error bars result from fluctuation of the solitons in the region where the distance is averaged.
}
\end{figure}

If the separation distance is disconnected to the fluid properties, only remains the driven-dissipative nature of the fluid. In order to better understand which parameters control the equilibrium distance, we numerically investigate the influence of the decay rate of the polaritons $\gamma_{\text{polariton}}$, plotted in figure \ref{fig:dsepgamma}. In the simulations, the speed of sound $c_s$ and the speed of the fluid $v_f$ are kept constant and correspond to the typical experimental ones; only the polariton decay rate changes. Figure \ref{fig:dsepgamma} clearly shows that the lifetime of the polariton plays a central role in fixing the equilibrium distance of the solitons: the more dissipation, the closer the solitons align to each other. Indeed the refilling of the inner region is only due to the polariton tunneling across the soliton, hence, the achievable polariton density in-between the solitons decreases as the polariton decay rate increases; the equilibrium between the density pressure and the soliton repulsion takes therefore place for a smaller separation distance.
In figure \ref{fig:dsepgamma}, the error bars also result from the small breathing, responsible for small oscillations of the solitons in the propagation range where the separation distance is averaged.
For our sample, the polariton lifetime is 10 ps, which corresponds to a linewidth of 0.07 meV. In figure \ref{fig:dsepgamma}, the blue dot illustrates the experimental result and conditions, i.e. the average equilibrium separation distance for the decay rate of our microcavity, in good agreement with the theory.

\section{Conclusion}

In conclusion, we implemented a new technique to imprint phase patterns on a quasi-resonantly pumped polariton fluid by engineering the driving field with an SLM. By imprinting the solitons and exploiting the optical bistability, we enhanced their propagation distance for more than one order of magnitude with respect to the previous works. This technique permits not only to artificially form dark solitons at will, but also  to study the solitons interaction: remarkably we have observed an unconventional binding mechanism, leading to the formation of a new kind of dark-soliton molecule in a local nonlinear medium. Moreover we demonstrated that the characteristic separation distance between the solitons of the molecule is governed by the driven-dissipative nature of the polariton fluid. 
This result opens unprecedented possibilities to control and manipulate collective excitations such as solitons and quantum vortices in quantum fluids of light.

\section*{Acknowledgements}
This work has received funding from the French ANR grant (C-FLigHT 138678, QFL) and from the European Union’s Horizon 2020 Research and Innovation Program under grant agreement No 820392 (PhoQuS). QG and AB thank the Institut Universitaire de France (IUF) for support.

\bibliography{LKB-bibs-bib_ImprSol.bib}

\end{document}